\documentclass[prl,twocolumn,superscriptaddress,10pt]{revtex4-1}
\usepackage{graphicx}
\usepackage{dcolumn}
\usepackage{color}
\usepackage{times}
\usepackage{bm}
\usepackage{amssymb}
\usepackage{amsmath}
\usepackage{epsfig}
\usepackage{epstopdf}
\usepackage{dsfont}
\usepackage{subfigure}
\usepackage{multirow}
\usepackage[colorlinks=true,citecolor=blue, linkcolor=blue]{hyperref}%

\begin{document}
	
	\title{\textbf{Experimental observation the Einstein-Podolsky-Rosen Steering based on the detection of entanglement}}
	
	\author{Huan Yang}
	\affiliation{School of Physics and Material Science, Anhui University, Hefei 230601, China}
	\affiliation{Institutes of Physical Science and Information Technology, Anhui University, Hefei 230601, China}
	\affiliation{Department of Experiment and Practical Training Management, West Anhui University, Lu’an 237012, China}
	
	\author{Zhi-Yong Ding}
	\affiliation{School of Physics and Material Science, Anhui University, Hefei 230601, China}
	\affiliation{School of Physics and Electronic Engineering, Fuyang Normal University, Fuyang 236037, China}
	\affiliation{Key Laboratory of Functional Materials and Devices for Informatics of Anhui Educational Institutions, Fuyang Normal University, Fuyang 236037, China}
	
	\author{Dong Wang}
	\affiliation{School of Physics and Material Science, Anhui University, Hefei 230601, China}
	\affiliation{CAS Key Laboratory of Quantum Information, University of Science and Technology of China, Hefei 230026, China}	
	
	\author{Hao Yuan}
	\affiliation{School of Physics and Material Science, Anhui University, Hefei 230601, China}
	\affiliation{CAS Key Laboratory of Quantum Information, University of Science and Technology of China, Hefei 230026, China}	
	\affiliation{Key Laboratory of Opto-Electronic Information Acquisition and Manipulation of Ministry of Education, Anhui University, Hefei 230601, China} 
	
	\author{Xue-Ke Song}
	\affiliation{School of Physics and Material Science, Anhui University, Hefei 230601, China}
	
	\author{Jie Yang}
	\affiliation{School of Physics and Material Science, Anhui University, Hefei 230601, China}
	
	\author{Chang-Jin Zhang}
	\affiliation{Institutes of Physical Science and Information Technology, Anhui University, Hefei 230601, China}
	
	\author{Liu Ye}
	\email[]{yeliu@ahu.edu.cn}
	\affiliation{School of Physics and Material Science, Anhui University, Hefei 230601, China}
	
	\begin{abstract}
		The Einstein-Podolsky-Rosen (EPR) steering is an intermediate quantum nonlocality between entanglement and Bell nonlocality, which plays an  important role in quantum information processing tasks. In the past few years, the investigations concerning EPR steering have been demonstrated in a series of experiments. However, these studies rely on the relevant steering inequalities and the choices of measurement settings. Here, we experimentally verify the EPR steering via entanglement detection without using any steering inequality and measurement setting. By constructing two new states from a two-qubit target state, we observe the EPR steering by detecting the entanglement of these new states. The results show that the entanglement of the newly constructed states can be regarded as a new kind of steering witness for target states. Compared to the results of Xiao \textit{et al}. [Phys. Rev. Lett. 118, 140404 (2017)], we find that the ability of detecting EPR steering in our scenario is stronger than two-setting projective measurements, which can observe more steerable states. Hence, our demonstrations can deepen the understanding of the connection between the EPR steering and entanglement.
	\end{abstract}

	\maketitle
	
	%\section{INTRODUCTION}
		\textit{Introduction.}--The investigations of Einstein-Podolsky-Rosen (EPR) steering, which was considered as a “spooky” action permissible under the rules of quantum mechanics, can be traced back to 1935 \cite{w01, w02}. Assuming two entangled parties shared by Alice and Bob in separated space, the EPR steering depicts a distinctive non-classical trait that Alice can immediately steer the Bob’s system by implementing a measurement on her system. Recently, Wiseman \textit{et al}. rigorously and operationally defined the EPR steering as an information-theoretic task \cite{w03}. The EPR steering exists in the situation that the joint probabilities of measurement outcomes cannot be described by constructing a local hidden variable-local hidden state (LHV-LHS) model. The hierarchy of quantum nonlocalities, namely, steerable states are a strict subset of the entangled states \cite{w04} and a superset of Bell nonlocal states \cite{w05, w06}, was also provided \cite{w03}. The EPR steering can thus be regarded as a entanglement witness, i.e., the entanglement must exist in a steerable state. In the past decade, the explorations concerning EPR steering have attracted increasing attention since it has many potential applications in quantum information processing, for example, one-sided device-independent quantum key distribution \cite{w07,w08,w09}, secure quantum teleportation \cite{w10,w11}, one-sided device-independent randomness generation \cite{w12}, and so on.
				
		One of the important features of the EPR steering is its inherent asymmetric characteristic differ from entanglement and Bell nonlocality \cite{w03,w13}, which has the unique advantage in several quantum information tasks. That is, there exist two-qubit entangled states in which the steerability from two directions are not equivalent (the ability of Alice to steer Bob is not equal to the ability of Bob to steer Alice). The EPR steering can be detected by the violation of various steering inequalities, including linear steering inequalities\cite{w14,w15}, inequalities based on entropic uncertainty relations \cite{w16,w17,w18,w19}, and steering criterion from geometric Bell-like inequality \cite{w20}. On the theory side, based on the choices of measurements, the EPR steering have been explored from projective measurements \cite{w13,w21} to positive operator-valued measures (POVMs) \cite{w22,w23,w24}, and also from continuous variable systems \cite{w25,w26} to discrete systems \cite{w13,w21,w23} .
						
		Experimentally, the EPR steering were investigated by a few efforts in the past few years, and many significant results were demonstrated. Saunders \textit{et al}. experimentally observed quantum steering for Bell local state via linear steering inequality \cite{w27}, and demonstrated that EPR steering occurs for mixed entangled states that are Bell local. Bennet \textit{et al}. experimentally certified that EPR steering can be rigorously implemented in the case of arbitrarily high losses \cite{w28}. The EPR steering of Gaussian states was verified by performing Gaussian measurements \cite{w29}, and then was extended to a multipartite system \cite{w30}. By using the steering radius, Sun \textit{et al}. \cite{w31} and Xiao \textit{et al}. \cite{w32} explored the EPR steering in different directions through two-setting and multisetting projective measurements, respectively. The results show that the phenomenon of EPR steering may disappear with the increase of measurement setting in experiment. Considering 16 measurements and a Werner state with a lossy channel at one side, Wollmann \textit{et al}. observed the EPR steering in the general setting of POVMs \cite{w24}. Without assumption concerning the experimental state or measurement, Tischler \textit{et al}. verified the asymmetry of EPR steering in the situation of a two-qubit state with loss \cite{w33}. Also, Li \textit{et al}. explored the EPR steering of two-qubit Werner state via the geometric Bell-like inequality \cite{w34}. However, these experimental efforts rely on the relevant steering inequalities and the choices of measurement settings, which make them difficult to avoid the locality loophole presented in EPR steering tests. In general, an entangled state can be verified by detecting EPR steering, and the converse is not always true \cite{w22}. The detection of EPR steering is also strictly harder than the detection of entanglement \cite{w27}. Therefore, an open question raises: can we realize the observation of EPR steering by detecting the entanglement in experiment? That is still lacking, and the study may provide a new way to test a difficultly detected quantum correlation by translating it into an easily detected problem.
		
		In this paper, we demonstrate the EPR steering in two directions via detecting entanglement in experiment, without using any steering inequality and measurement setting. To be specific, we experimentally construct two new states from a two-qubit target state on the basis of the recent theories of Das \textit{et al}. \cite{w35} and Chen \textit{et al}. \cite{w36}, and observe the EPR steering by detecting the entanglements of the new states for the first time. The results verify that the entanglement of newly constructed states can be considered as a new kind of EPR steering witness for target states in experiment. These experimental results can deepen our understanding of the connection between steering and entanglement. We further compare our results with the ones of Xiao \textit{et al}. \cite{w32}, and it is shown that the ability of detecting EPR steering in our scenario is stronger than two-setting projective measurements. It has a wider region for the observation of steerable states.
			
	%\section{Observing EPR steering through entanglement detection}
		\textit{Observing EPR steering via entanglement detection.}--Considering a bipartite quantum state ${\rho _{AB}}$ shared by Alice and Bob, and the possible choices of measurements are \textit{A} (measurement operators denoted by  $M_a^A$ ) and \textit{B} (measurement operators denoted by $M_b^B$) for Alice and Bob, respectively. Assuming that Alice obtains the measurement outcome \textit{a} by carrying out the measurement of \textit{A} on her subsystem, Bob obtains corresponding outcome \textit{b} by implementing the measurement \textit{B} on his subsystem. If and only if the corresponding joint probability distribution of the outcomes cannot be expressed by $P(a,b|A,B,{\rho _{AB}}) = \sum\nolimits_\lambda  {P(\lambda )P(a|A,\lambda ){P_Q}(b|B,{\rho _\lambda })}$, and then the ${\rho _{AB}}$ is steerable from Alice to Bob \cite{w03}. Here, $P(\lambda )$ is the probability distribution over the LHVs satisfying $\sum\nolimits_\lambda  {p(\lambda )}  = 1$. $P(a|A,\lambda )$ indicates a general probability distribution, and ${P_Q}(b|B,{\rho _\lambda }) = Tr[{\rho _\lambda }M_b^B]$ is the probability of outcome  \textit{b} performed measurement \textit{B} on the LHS ${\rho _\lambda }$. In other words, the ${\rho _{AB}}$ can realize steering from Alice to Bob if and only if there is no LHV-LHS model descripted by the joint probability distribution for arbitrary measurements implemented by Alice and Bob.		
		
		Any bipartite separable state is defined as a convex mixture of product states, namely, ${\rho _{AB}} = \sum\nolimits_\lambda  {p(\lambda )\rho _\lambda ^A \otimes \rho _\lambda ^B}$. The state ${\rho _{AB}}$ is entangled if and only if the joint probability distribution cannot be represented by $P(a,b|A,B,{\rho _{AB}}) = \sum\nolimits_\lambda  {P(\lambda ){P_Q}(a|A,\rho _\lambda ^A){P_Q}(b|B,\rho _\lambda ^B)}$, where ${P_Q}(a|A,\rho _\lambda ^A) = Tr[\rho _\lambda ^AM_a^A]$ and ${P_Q}(b|B,\rho _\lambda ^B) = Tr[\rho _\lambda ^BM_b^B]$. That is, the ${\rho _{AB}}$ is an entangled state if and only if there is no LHS-LHS model descripted by this joint probability distribution for arbitrary measurements performed by Alice and Bob. Recently, Das \textit{et al}. \cite{w35} and Chen \textit{et al}. \cite{w36} proposed a sufficient criteria of detecting the EPR steering for bipartite states by the entanglement detection of newly constructed states. In theory, for any two-qubit state ${\rho _{AB}}$ shared by Alice and Bob, we can construct two new states as $\tau _{AB}^1$  and $\tau _{AB}^2$,		
		\begin{equation}\label{g01}
			\tau _{{\rm{AB}}}^1 = {\mu _1}{\rho _{AB}} + (1 - {\mu _1})\tilde \rho _{AB}^1,
		\end{equation}
		and 
		\begin{equation}\label{g02}
		\tau _{{\rm{AB}}}^2 = {\mu _2}{\rho _{AB}} + (1 - {\mu _2})\tilde \rho _{AB}^2.
		\end{equation}	
		Here, $\tilde \rho _{AB}^1 = {\rho _A} \otimes\mathbb{I}/2$,  ${\rho _A} = {\rm{T}}{{\rm{r}}_B}\left[ {{\rho _{AB}}} \right]$, $\tilde \rho _{AB}^2 = \mathbb{I}/2 \otimes {\rho _B}$, ${\rho _B} = {\rm{T}}{{\rm{r}}_A}\left[ {{\rho _{AB}}} \right]$,${\mu _1} \in [0,{\rm{ }}1/\sqrt 3 ]$, and ${\mu _2} \in [0,{\rm{ }}1/\sqrt 3 ]$. If $\tau _{AB}^1$ is an entangled state, then the state ${\rho _{AB}}$ is steerable from Bob to Alice. In addition, if $\tau _{AB}^2$ is entangled, then the steering task from Alice to Bob can be realized.
		
		In order to observe the EPR steering in experiment, the steerable states need to be prepared in our photon-polarization-qubit system. We focus our attention on a family of two-qubit target states \cite{w37}:
		\begin{equation}\label{g03}
		{\rho _{AB}}(\alpha ,\theta ) = \alpha \left| {\psi (\theta )} \right\rangle \left\langle {\psi (\theta )} \right| + (1 - \alpha )\mathbb{I}/2 \otimes {\rho _B}
		\end{equation} 
		with $\left| {\psi (\theta )} \right\rangle  = \cos (\theta )\left| {HH} \right\rangle  + \sin (\theta )\left| {VV} \right\rangle $ and ${\rho _B} = {\rm{T}}{{\rm{r}}_A}\left[ {\left| {\psi (\theta )} \right\rangle \left\langle {\psi (\theta )} \right|} \right]$. Here, $0 \le \alpha  \le 1$ and $0 \le \theta  \le \pi /{\rm{4}}$. The horizontally and vertically polarized components are represented by \textit{H} and \textit{V}, respectively. It is shown that the state is steerable from Alice to Bob if $\alpha  > 1/2$. Moreover, it is demonstrated that Bob cannot steer Alice for ${\cos ^2}(2\theta ) \ge (2\alpha  - 1)/(2 - \alpha ){\alpha ^3}$ by employing the uniform distribution as an ansatz for the LHS ensemble \cite{w37}. According to Refs. \cite{w35} and \cite{w36}, the newly constructed states $\tau _{AB}^1(\alpha ,\theta )$ and $\tau _{AB}^2(\alpha ,\theta )$ for ${\rho _{AB}}(\alpha ,\theta )$ can be given by
		\begin{equation}\label{g04}
		\tau _{AB}^1(\alpha ,\theta ) = {\mu _1}{\rho _{AB}}(\alpha ,\theta ) + (1 - {\mu _1})\tilde \rho _{AB}^1(\alpha ,\theta ),
		\end{equation} 
		and 	
		\begin{equation}\label{g05}
		\tau _{AB}^2(\alpha ,\theta ) = {\mu _2}{\rho _{AB}}(\alpha ,\theta ) + (1 - {\mu _2})\tilde \rho _{AB}^2(\alpha ,\theta ),
		\end{equation}		
		where $\tilde \rho _{AB}^1(\alpha ,\theta ) = {\rho _A}(\alpha ,\theta ) \otimes \mathbb{I}/2$ with ${\rho _A}(\alpha ,\theta ) = {\rm{T}}{{\rm{r}}_B}\left[ {{\rho _{AB}}(\alpha ,\theta )} \right]$, $\tilde \rho _{AB}^2(\alpha ,\theta ) = \mathbb{I}/2 \otimes {\rho _B}(\alpha ,\theta )$ with ${\rho _B}(\alpha ,\theta ) = {\rm{T}}{{\rm{r}}_A}\left[ {{\rho _{AB}}(\alpha ,\theta )} \right]$. It is known that the entanglement of two-qubit states can be effectively identified by the concurrence \cite{w38}. Considering a two-qubit state ${\rho _{AB}}$, the concurrence is written as $C({\rho _{AB}}) = \max \{ 0,\sqrt {{\lambda _1}}  - \sqrt {{\lambda _2}}  - \sqrt {{\lambda _3}}  - \sqrt {{\lambda _4}} \} $, where ${\lambda _i}(i = 1,2,3,4)$ are the eigenvalues with decreasing order of the matrix ${\rho _{AB}}({\sigma _y} \otimes {\sigma _y})\rho _{AB}^*({\sigma _y} \otimes {\sigma _y})$. The variable $\rho _{AB}^*$ represents the complex conjugate of ${\rho _{AB}}$ in the fixed basis $\{ \left| {00} \right\rangle ,\left| {01} \right\rangle ,\left| {10} \right\rangle ,\left| {11} \right\rangle \} $. By some calculations, the condition for the case that $\tau _{AB}^1(\alpha ,\theta )$ is an entangled state (Bob can steer Alice) can be given by
		\begin{equation}\label{g06}
		\frac{{\sqrt 3  - \sqrt 3 \cos (4\theta ) - 2\sqrt {7 - 4\cos (4\theta ) + {{\cos }^2}(4\theta )} }}{{\cos (4\theta ) - 5}} < \alpha  \le 1
		\end{equation}		
		with $0 < \theta  \le \pi /{\rm{4}}$. The condition of $\tau _{AB}^2(\alpha ,\theta )$ being an entangled state (Alice can steer Bob) can be written as $1/\sqrt 3  < \alpha  \le 1$ and $0 < \theta  \le \pi /{\rm{4}}$. Hence, to make sure that $\tau _{AB}^2(\alpha ,\theta )$ is an entangled state and $\tau _{AB}^1(\alpha ,\theta )$ is a separable state, it should satisfy the following condition
		\begin{equation}\label{g07}
		\frac{1}{{\sqrt 3 }} < \alpha  \le \frac{{\sqrt 3  - \sqrt 3 \cos (4\theta ) - 2\sqrt {7 - 4\cos (4\theta ) + {{\cos }^2}(4\theta )} }}{{\cos (4\theta ) - 5}}
		\end{equation}
		with $0 < \theta  < \pi /{\rm{4}}$.	The detailed calculations are given in the Supplemental Material.			
		
	%\section{EXPERIMENTAL IMPLEMENTATION AND RESULTS}
		\textit{Experimental implementation and results.}--Fig. 1 provides the schematic diagram of our experimental setup. The setup consists of three modules: (a) state preparation, (b) new state construction, and (c) entanglement detection. To be specific, a tunable diode laser emits a center wavelength of 405nm and high-power (130mW) continuous laser beam, which passes through the polarization beam splitter (PBS). Subsequently, the transmitted beam is passed through a 405nm half-wave plate (HWP1) and focused on two type-I $\beta $-barium borate (BBO) crystals $\left( {6.0 \times 6.0 \times 0.5{\rm{mm}}} \right)$. The state $\left| {\psi (\theta )} \right\rangle  = \cos (\theta )\left| {HH} \right\rangle  + \sin (\theta )\left| {VV} \right\rangle $ shared by a pair of entangled photons $\left( {\lambda  = 810{\rm{nm}}} \right)$ are generated via spontaneous parametric down-conversion (SPDC) \cite{w39}. We can control the state parameter $\theta $ by adjusting the rotation angle of HWP1. We insert an unbalanced interferometer (UI0) in the path of A, and the 50/50 beam splitter (BS) in the UI0 separate the photon into two paths (represented by ${p_1}$ and ${p_2}$). In the path ${p_1}$, the state of photons remains unchanged. The path ${p_2}$ is composed by a HWP2 with ${22.5^{\rm{o}}}$ and three $321\lambda $ yttrium orthovanadate ${\rm{(YV}}{{\rm{O}}_4})$ crystal, which can completely destroy the coherence. The state shared by the photons of path ${p_2}$ and path B in module (a) is transformed into an incoherent state in the basis of eigenvectors of Pauli matrix ${\sigma _z}$, and the diagonal elements are ${\cos ^2}\theta /2$, ${\sin ^2}\theta /2$, ${\cos ^2}\theta /2$, and ${\sin ^2}\theta /2$, respectively. The two-qubit states ${\rho _{AB}}(\alpha ,\theta )$ can be prepared by combining the paths ${p_1}$ and ${p_2}$ into one, and the state parameter $\alpha $ can be changed conveniently by using attenuator (ATT) in the UI0. 30 states ${\rho _{AB}}(\alpha ,\theta )$ are prepared to carry out the EPR steering test in our experiment. Their distributions are denoted by the red, blue and green solid circles in Fig. 2, and the distributions of theoretical steerable states ${\rho _{AB}}(\alpha ,\theta )$ are also displayed by different color regions. The red region represents the states ${\rho _{AB}}(\alpha ,\theta )$ for which $\tau _{AB}^1(\alpha ,\theta )$ and $\tau _{AB}^2(\alpha ,\theta )$ are not entangled states. The resluts imply that the EPR steering in both directions cannot be observed through entanglement detection in theory. The blue region, which is depicted by Eq. (7), represents the states ${\rho _{AB}}(\alpha ,\theta )$ for which only $\tau _{AB}^2(\alpha ,\theta )$ is entangled state. That is to say, only the EPR steering from Alice to Bob can be observed via entanglement detection, whether Bob can steer Alice cannot be determined. The green region represents the case that both $\tau _{AB}^1(\alpha ,\theta )$ and $\tau _{AB}^2(\alpha ,\theta )$ are entangled states, and the steering task succeeds in both directions. In our method, we observe the EPR steering by detecting the entanglement of these new states $\tau _{AB}^1(\alpha ,\theta )$ and $\tau _{AB}^2(\alpha ,\theta )$, without using any steering inequality and measurement setting. This is distinct from the ones in Ref. \cite{w32}, which investigated the EPR steering of ${\rho _{AB}}(\alpha ,\theta )$ by using the multimeasurement settings and the steering radius. Note that, Fig. 2 also provides the results of Ref. \cite{w32}. Specifically, in the case of two-measurement settings, the states below the red solid line cannot realize steering task. The states between the red solid line and dashed curve mean that Alice can steer Bob, but Bob cannot steer Alice. The states above the red dashed curve can achieve the steering task in both directions. The blue solid line and dashed curve are the results in three-measurement settings. The region below the blue solid line indicates that the steering task fails in both directions. The states in the region between the blue solid line and dashed curve signify that only Alice can steer Bob, and the region above the blue dashed curve represents that the states are both-way steerable. In comparison with the two-measurement settings, the scenario of entanglement detection can capture more steerable states, while it is approximately equal to the three-measurement settings in the detecting of EPR steering from Alice to Bob. We also depict the steering region in infinite-measurement settings, which is bounded by the black solid curve. The states below the curve can only realize steering from Alice to Bob \cite{w37}.
						
		\begin{figure}
			\centering
			\includegraphics[width=8cm]{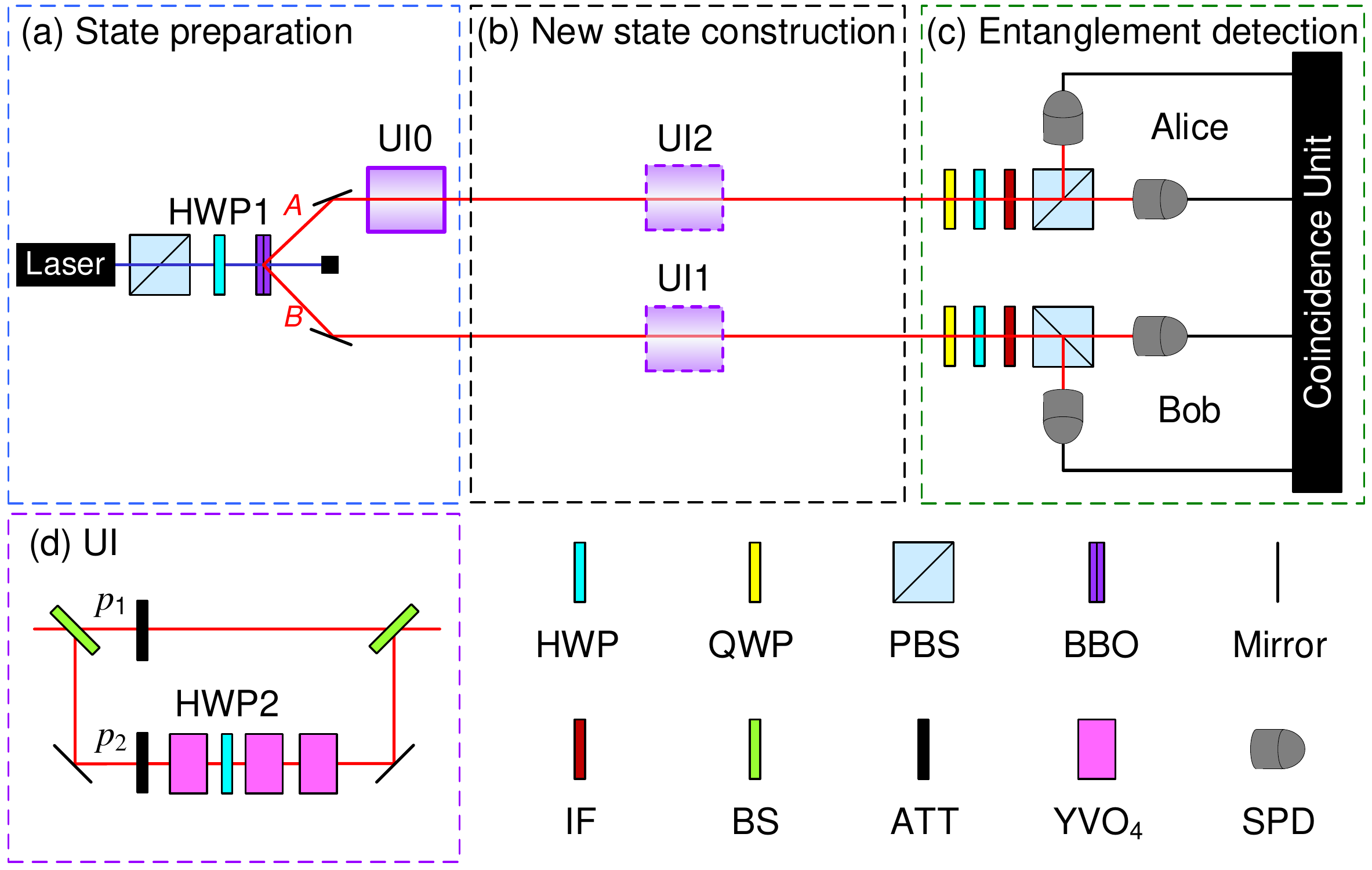}\\
			\caption{Experimental setup. The setup includes three modules: (a) state preparation, (b) new state construction, and (c) entanglement detection. In brief, a family of two-qubit states ${\rho _{AB}}(\alpha ,\theta )$ are prepared by the procession of a spontaneous parametric down-conversion and the unbalanced interferometer (UI0) in module (a). In module (b), two new states $\tau _{AB}^1(\alpha ,\theta )$ and $\tau _{AB}^2(\alpha ,\theta )$ from any ${\rho _{AB}}(\alpha ,\theta )$ are constructed. If we only let the unbalanced interferometer (UI1) influence on the path B of module (b), and remain the path A unchanged, the state $\tau _{AB}^1(\alpha ,\theta )$ can be prepared. On the contrary, if we let the unbalanced interferometer (UI2) influence on the path A of module (b), and remain the path B unchanged, the state $\tau _{AB}^2(\alpha ,\theta )$ can be generated. Module (c) is used to realize the entanglement detection of the new states. Abbreviations: HWP, half-wave plate; QWP, quarter-wave plate; PBS, polarizing beam splitter; BBO, type-I $\beta$ -barium borate; IF: interference filter; BS: beam splitter; ATT: attenuator; ${\rm{YV}}{{\rm{O}}_4}$: yttrium orthovanadate crystal; SPD: single photon detector.}\label{Fig1}
		\end{figure}

		\begin{figure}
			\centering
			\includegraphics[width=8cm]{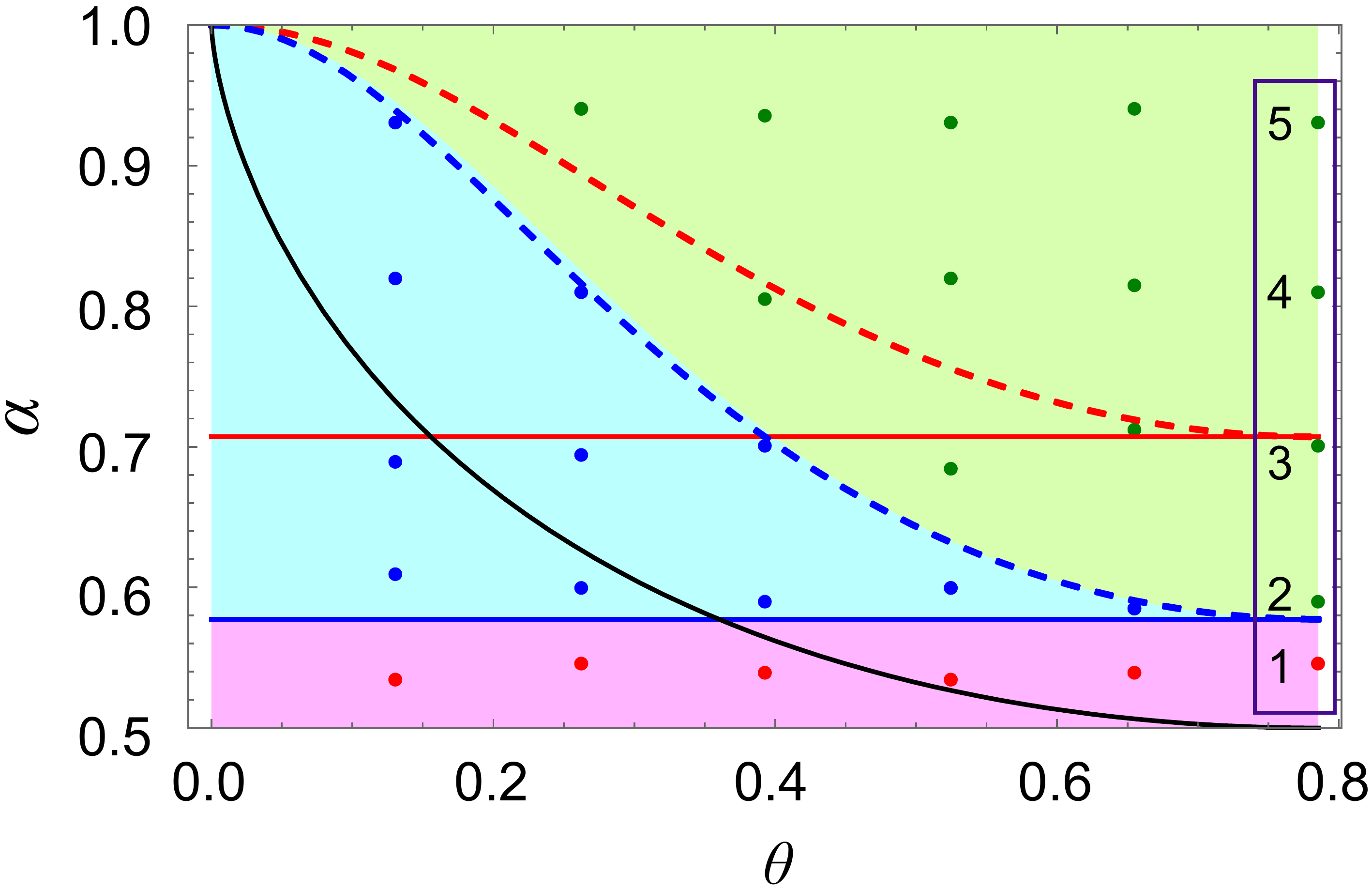}\\
			\caption{Distributions of states ${\rho _{AB}}(\alpha ,\theta )$  both in our experiment and in different scenarios. In the theoretical scenario of observing the EPR steering through entanglement detection, the red area represents the states that the EPR steering in both directions cannot be verified. The blue area denotes the states that only the EPR steering from Alice to Bob can be observed in this theory. The green area represents the states for which Alice and Bob can steer each other. The red, blue, and green solid circles in the corresponding regions are the experimentally prepared states. The red solid line and dashed curve represent the results in the scenario of two-measurement settings \cite{w32}. The region below the red solid line (above the red dashed curve) represents the states are unsteerable (both-way steerable) states. The states between the red solid line and dashed curve can only realize the EPR steering form Alice to Bob. In the case of three-measurement settings \cite{w32}, the states below the blue solid line (above the blue dashed curve) are unsteerable (both-way steerable). The region between the blue solid line and dashed curve represents that only the steering task from Alice to Bob can be achieved. The region between the \textit{x} axis and black solid curve denotes that the states are steerable from Alice to Bob, and the states above the black curve are both-way steerable states in the scenario of infinite-measurement settings \cite{w37}.}\label{Fig2}		
		\end{figure}

		Next let us construct new states $\tau _{AB}^1(\alpha ,\theta )$ and $\tau _{AB}^2(\alpha ,\theta )$ in our all-optical setup, as shown in module (b) of Fig. 1. To construct $\tau _{AB}^1(\alpha ,\theta )$, we block path ${p_2}$ in the UI2, and only let the unbalanced interferometer (UI1) influence on the path B. In other words, we let the photon of path B in module (b) send to UI1 and then send to Bob. The photon of path A is sent to Alice directly. The photon state of path  ${p_1}$ in UI1 remains unchanged. The state shared by path ${p_2}$ in UI1 and path A is dephased into an incoherent state, and the weights in bases of $\left| {HH} \right\rangle \left\langle {HH} \right|$, $\left| {HV} \right\rangle \left\langle {HV} \right|$, $\left| {VH} \right\rangle \left\langle {VH} \right|$, and $\left| {VV} \right\rangle \left\langle {VV} \right|$ are $(1 + \alpha \cos 2\theta )/4$, $(1 + \alpha \cos 2\theta )/4$, $(1 - \alpha \cos 2\theta )/4$, and $(1 - \alpha \cos 2\theta )/4$, respectively. The $\tau _{AB}^1(\alpha ,\theta )$ can be constructed by combining these two paths  ${p_1}$ and ${p_2}$ (in the UI1) into one in the experiment. The state parameter ${\mu _1}$ is set to $1/\sqrt 3  \approx 0.58$ by adjusting ATT in the UI1. The detailed method for determining the ${\mu _1}$ is shown in the Supplemental Material. Similarly, in order to construct $\tau _{AB}^2(\alpha ,\theta )$, we need block path ${p_2}$ in the UI1, and let the UI2 influence the path A. That is, the photon of path B in module (b) is sent to Bob directly, and the photon of path A is send to UI2 and then send to Alice. Two-photon state shared by the photons in the path ${p_2}$ of UI2 and path B is transformed into an incoherent state, the diagonal elements are ${\cos ^2}\theta /2$, ${\sin ^2}\theta /2$, ${\cos ^2}\theta /2$, and ${\sin ^2}\theta /2$, respectively. The state $\tau _{AB}^2(\alpha ,\theta )$ can be constructed by mixing the paths ${p_1}$ and ${p_2}$ of UI2 into one. The state parameter ${\mu _2}$ is also set to $1/\sqrt 3  \approx 0.58$ by using ATT in the UI2. Module (c) is used to realize the detection of entanglement by performing a quantum state tomography process \cite{w40}. The fidelity of $\tau _{AB}^1(\alpha ,\theta )$ and $\tau _{AB}^2(\alpha ,\theta )$ are calculated by $F(\tau ,{\tau _0}) \equiv {\rm{Tr}}\sqrt {\sqrt \tau  {\tau _0}\sqrt \tau  } $ \cite{w41}, where $\tau $ and ${\tau _0}$ are the experimental and theoretical density matrices, respectively. In our experiment, 30 new states $\tau _{AB}^1(\alpha ,\theta )$ and 30 $\tau _{AB}^2(\alpha ,\theta )$ are constructed, and the fidelities of all these states are beyond 0.9873.

		\begin{figure}
			\centering
			\includegraphics[width=8cm]{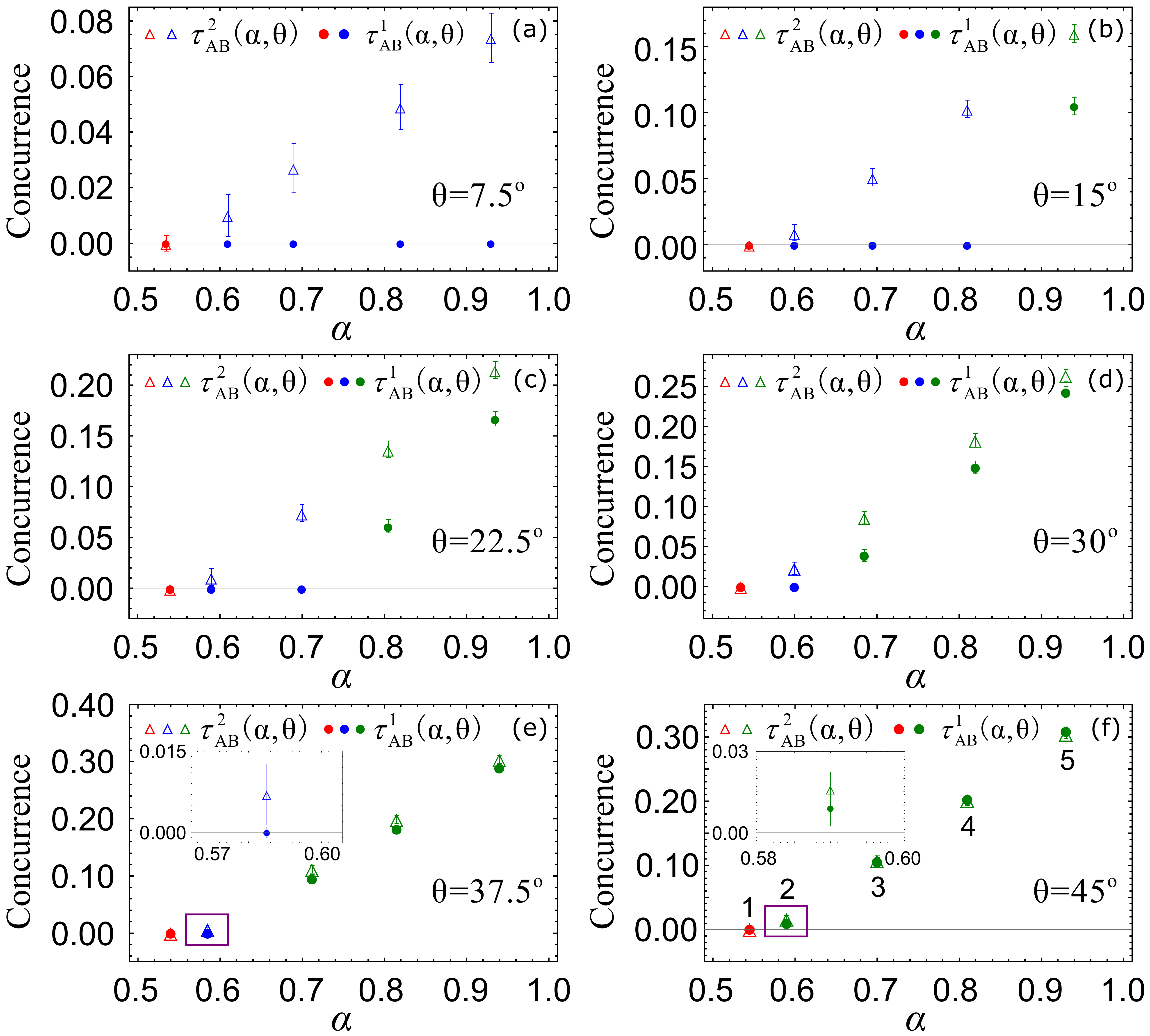}\\
			\caption{The experimental results. The solid circles and the hollow triangles represent the experimental results of $C(\tau _{AB}^1(\alpha ,\theta ))$ and $C(\tau _{AB}^2(\alpha ,\theta ))$, respectively. The red solid 	circles and hollow triangles correspond to the results that the $\tau _{AB}^1(\alpha ,\theta )$ and $\tau _{AB}^2(\alpha ,\theta )$ are all separated, i.e., the EPR steering of ${\rho _{AB}}(\alpha ,\theta )$ cannot be verified through entanglement detection. The blue solid circles and hollow triangles correspond to the case for which only the $\tau _{AB}^2(\alpha ,\theta )$ is entangled, namely, only the steering from Alice to steer Bob can be observed via entanglement detection. The green solid circles and hollow triangles correspond to the results that $\tau _{AB}^1(\alpha ,\theta )$ and $\tau _{AB}^2(\alpha ,\theta )$ are all entangled, i.e., Alice and Bob can steer each other. Some of error bars are very small and not shown.}\label{Fig3}	
		\end{figure}

		\begin{figure}
			\centering
			\includegraphics[width=8cm]{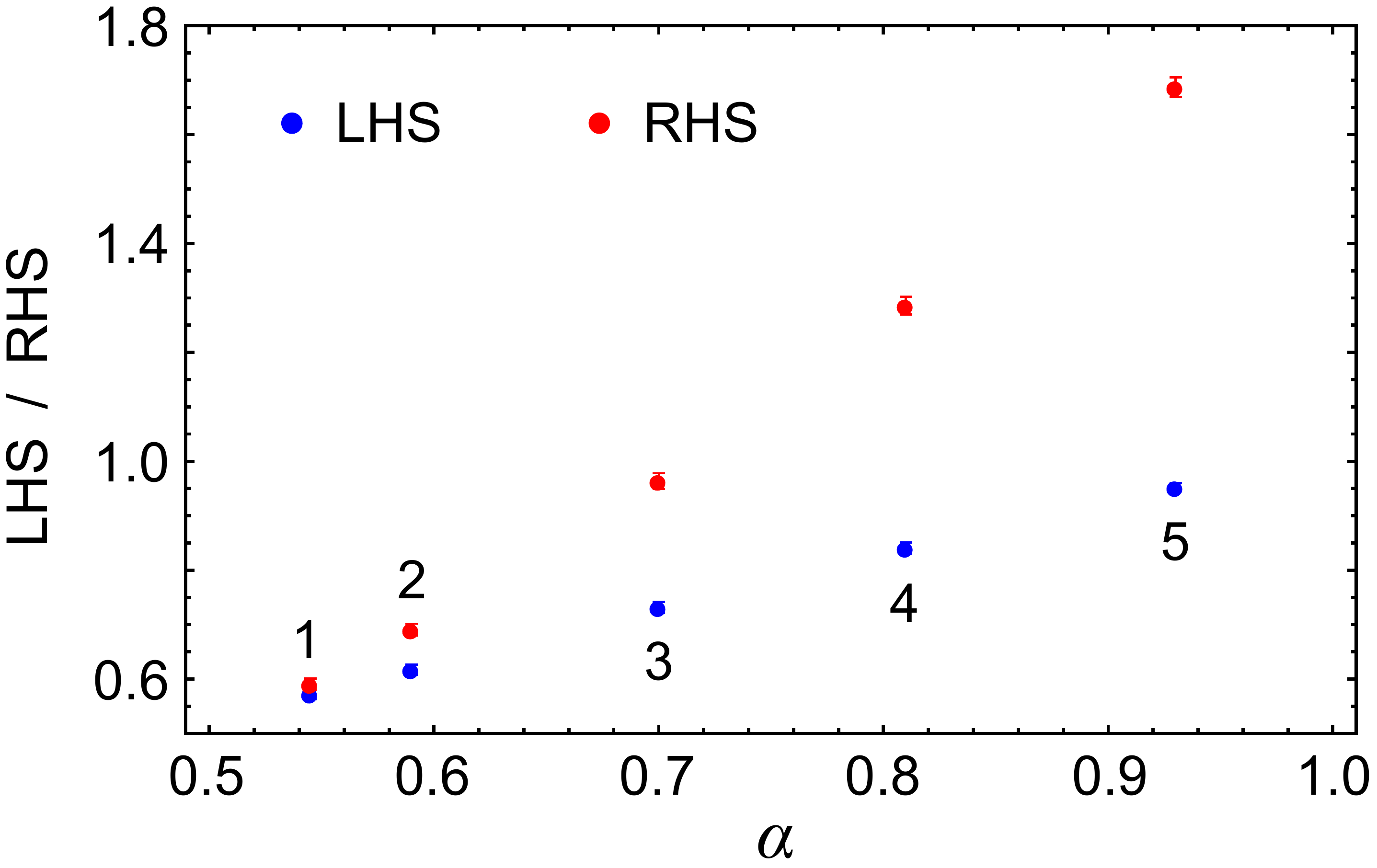}\\
			\caption{The experimental results for ${\rho _{AB}}(\alpha ,\pi /4)$ through the geometric Bell-like inequality. The blue and red solid circles denote the left-hand side (LHS) and the right-hand side (RHS) of the geometric Bell-like inequality, respectively. The experimental datas labelled by 1, 2, 3, 4, and 5 correspond to the test results of state labelled by 1, 2, 3, 4, and 5 in Fig. 2, respectively. Some of error bars are too small to display.}\label{Fig4}		
		\end{figure}

		In the following, we observe the EPR steering by detecting the entanglement of these 60 new states. The results of experiment are shown in Fig. 3, and the insets in Figs. 3(e) and (f) are the magnification of the regions in the purple panes. The solid circles and the hollow triangles represent the experimental results of $C(\tau _{AB}^1(\alpha ,\theta ))$ and $C(\tau _{AB}^2(\alpha ,\theta ))$, respectively, which are calculated according to the density matrices $\tau _{AB}^1(\alpha ,\theta )$ and $\tau _{AB}^2(\alpha ,\theta )$ obtained by tomography. Based on the standard deviation from the statistical variation of the photon counts, which are assumed to follow Poisson distribution, all error bars are estimated in the experiment. Note that some of error bars are very small and not displayed in the Fig. 3. In order to make the experimental results correspond to the prepared states  in Fig. 2, we use solid circles and hollow triangles with different colors in Fig. 3 to represent the corresponding experimental results. As seen from Fig. 3, the six red solid circles and six red hollow triangles display that the experimental states $\tau _{AB}^1(\alpha ,\theta )$ and $\tau _{AB}^2(\alpha ,\theta )$ are all separable. The results certify that the EPR steering of six prepared states ${\rho _{AB}}(\alpha ,\theta )$  (represented by six red solid circles in Fig. 2) cannot be observed via the detection of entanglement. The eleven blue solid circles and eleven blue hollow triangles in Fig. 3 represent that the $\tau _{AB}^1(\alpha ,\theta )$ are separable states and the $\tau _{AB}^2(\alpha ,\theta )$ are entangled states, respectively. The experimental results can help us to identify that the eleven prepared states ${\rho _{AB}}(\alpha ,\theta )$  (represented by eleven blue solid circles in Fig. 2) can realize the EPR steering from Alice to Bob, and whether the EPR steering from Bob to Alice cannot be decided in this scenario. One can see from Fig. 3 that the thirteen green solid circles and thirteen green hollow triangles indicate that the $\tau _{AB}^1(\alpha ,\theta )$ and $\tau _{AB}^2(\alpha ,\theta )$ are all entangled states in experiment. The results verify that the thirteen prepared states ${\rho _{AB}}(\alpha ,\theta )$ (represented by thirteen green solid circles in Fig. 2) are both-way steering states, in which the steering task succeeds in both directions (Alice and Bob can steer each other). Hence, the experimental results show good agreement with theoretical ones in Fig. 2, and the EPR steering of two-qubit target states ${\rho _{AB}}(\alpha ,\theta )$ can be observed through the entanglement detection of these new states. As shown in Fig. 2, some steerable states (the states between the red and blue solid line), which can be demonstrated in our scenario, cannot be verified in the case of two-setting projective measurements. Nearly all steerable states, which can be observed in the case of three-setting projective measurements, can also witnessed by detecting entanglement. This means that the ability of detecting EPR steering in our strategy is stronger than two-measurement settings, and can compete with the three-measurement settings.
		
		In order to further verify the effectiveness of our scenario, we take the EPR steering of ${\rho _{AB}}(\alpha ,\pi /4)$  (labelled by 1, 2, 3, 4 and 5 in Fig. 2) by using the geometric Bell-like inequality as an example, which is denoted by $\mathop {{\rm{max}}}\limits_{\mathord{\buildrel{\lower3pt\hbox{$\scriptscriptstyle\rightharpoonup$}} 
					\over m} ,\mathord{\buildrel{\lower3pt\hbox{$\scriptscriptstyle\rightharpoonup$}} 
					\over n} } {E_Q}(\mathord{\buildrel{\lower3pt\hbox{$\scriptscriptstyle\rightharpoonup$}} 
				\over m} ,\mathord{\buildrel{\lower3pt\hbox{$\scriptscriptstyle\rightharpoonup$}} 
				\over n} ) \ge 2(\sum\nolimits_{i,j = 1}^3 {T_{ij}^2} )/3$ with ${E_Q}(\mathord{\buildrel{\lower3pt\hbox{$\scriptscriptstyle\rightharpoonup$}} 
				\over m} ,\mathord{\buildrel{\lower3pt\hbox{$\scriptscriptstyle\rightharpoonup$}} 
				\over n} ) = \sum\nolimits_{i,j = 1}^3 {{T_{ij}}} {m_i}{n_j}$ \cite{w20}. Here, $\mathord{\buildrel{\lower3pt\hbox{$\scriptscriptstyle\rightharpoonup$}} 
				\over m}  \cdot {\mathord{\buildrel{\lower3pt\hbox{$\scriptscriptstyle\rightharpoonup$}} 
					\over \sigma } ^A}$ and $\mathord{\buildrel{\lower3pt\hbox{$\scriptscriptstyle\rightharpoonup$}} 
				\over n}  \cdot {\mathord{\buildrel{\lower3pt\hbox{$\scriptscriptstyle\rightharpoonup$}} 
				\over \sigma } ^B}$ are projective measurements implemented by Alice and Bob, respectively. ${\mathord{\buildrel{\lower3pt\hbox{$\scriptscriptstyle\rightharpoonup$}} 
				\over \sigma } ^A}$ and ${\mathord{\buildrel{\lower3pt\hbox{$\scriptscriptstyle\rightharpoonup$}} 
				\over \sigma } ^B}$ are Pauli matrices, and ${T_{ij}} = Tr[{\rho _{AB}}(\sigma _i^A \otimes \sigma _j^B)]$ is the matrix element of the spin correlation matrix. We use the LHS and RHS to represent the left-hand side and the right-hand side of the geometric Bell-like inequality, respectively. Note that, the state ${\rho _{AB}}(\alpha ,\pi /4)$ corresponds to the Werner state \cite{w42}. It is well known that the state is steerable from Alice to Bob in the case of $\alpha  > 1/2$ , and Bob can steer Alice for $\alpha  > 1/2$ \cite{w03,w43}. Based on the quantum state tomography, the experimental LHS and RHS are calculated according to the density matrix of ${\rho _{AB}}(\alpha ,\pi /4)$ (the fidelities of all these states are higher than 0.9893). The experimental results are shown in Fig. 4. The both-way steering of the states labelled by 2, 3, 4, and 5 in Fig. 2 can be observed in our scenario (see Fig. 3 (f)), and these states also violate the geometric Bell-like inequality (the data of LHS are less than the ones of RHS, see Fig. 4). The results further verify that the entanglement detection of newly constructed state can effective witness the EPR steering in experiment. Meanwhile, it is worth noting that the EPR steering of the state labelled by 1 in Fig. 2 can be observed by employing the geometric Bell-like inequality, however, cannot be observed through our scenario. The reason is that the ability of geometric Bell-like inequality for detecting the EPR steering of Werner state (for  $\alpha  > 1/2$ \cite{w20,w34}) is stronger than the scenario of entanglement detection (for $\alpha  > 1/\sqrt 3 $).

	%\section{CONCLUSIONS}	
		\textit{Conclusions.}--In this paper, based on 30 prepared two-qubit states ${\rho _{AB}}(\alpha ,\theta )$, 60 new states $\tau _{AB}^1(\alpha ,\theta )$ and $\tau _{AB}^2(\alpha ,\theta )$ are constructed in the experiment. The EPR steering of ${\rho _{AB}}(\alpha ,\theta )$ are experimentally observed by detecting the entanglement of these newly constructed states, and any steering inequality and measurement setting are not used in the process of demonstration. Our results verify that the steering from Bob to Alice can be witnessed by the entanglement of $\tau _{AB}^1(\alpha ,\theta )$, and the steering from Alice to Bob can be witnessed by the entanglement of $\tau _{AB}^2(\alpha ,\theta )$. Hence, the entanglement of newly constructed state provides a new way to witness the EPR steering in experiment. The ability of testing the EPR steering in our scenario is stronger than the two-setting projective measurements, and more steerable states can be observed by detecting the entanglement of the constructed states. Consequently, our work demonstrates that one can effectively certify EPR steering by translating it into an easily certified quantum nonlocality (i.e., the entanglement) in experiment, and it is also potentially used to understand the relation between steering and entanglement in quantum information tasks.
		
	%\section*{ACKNOWLEDGMENTS}	
		This work was supported by the National Science Foundation of China (Grant Nos. 11575001, 61601002 and 11605028), the Program for Excellent Talents in University of Anhui Province of China (Grant No. gxyq2018059), the Natural Science Research Project of Education Department of Anhui Province of China (Grant No. KJ2018A0343, KJ2017A406 and KJ2017A401), the Open Foundation for CAS Key Laboratory of Quantum Information under Grant Nos. KQI201801 and KQI201804, the Key Program of Excellent Youth Talent Project of the Education Department of Anhui Province of China under Grant No. gxyqZD2018065.
		
		Huan Yang and Zhi-Yong Ding contributed equally to this work.

	\bibliographystyle{plain}
	
\end{document}